\documentclass[a4paper,floatfix,amsmath,amssymb,prb,twocolumn,showkeys,showpacs]{revtex4}
\usepackage{amsmath}
\usepackage{graphicx}
\usepackage{epsf}
\usepackage{dcolumn}
\usepackage{bm}
\usepackage{color}
\usepackage{caption}
\usepackage{subfig}
\begin{document}
\newcommand{\newc}{\newcommand}
\newc{\mbf}{\mathbf}
\newc{\boma}{\boldmath}
\newc{\beq}{\begin{equation}}
\newc{\eeq}{\end{equation}}
\newc{\beqar}{\begin{eqnarray}}
\newc{\eeqar}{\end{eqnarray}}
\newc{\beqa}{\begin{eqnarray*}}
\newc{\eeqa}{\end{eqnarray*}}
\newc{\bd}{\begin{displaymath}}
\newc{\ed}{\end{displaymath}}
\title{Revisiting Surface Diffusion in Random Deposition}
\author{ Baisakhi Mal} 
\affiliation{Department of Physics, Jadavpur University, Calcutta
700 032, India.}
\author{Subhankar Ray}
\email{sray@phys.jdvu.ac.in}
\affiliation{Department of Physics, Jadavpur University, Calcutta
700 032, India.}
\author{J. Shamanna}
\email{jlsphy@caluniv.ac.in}
\affiliation{Physics Department, University of Calcutta, Calcutta 700 009, India.}

\begin{abstract}
An investigation of the effect of surface diffusion in random
deposition model is made by analytical methods and reasoning.
For any given site, 
the extent to which a particle can diffuse is decided by 
the morphology in the immediate neighbourhood of the site. An 
analytical expression is derived to calculate the probability of
a particle at any chosen site to diffuse to a given length,
from first principles.
Using the method, the probabilities for different diffusion lengths
are calculated and their dependence on system size and the
number of deposited layers is studied. Numerical simulation of 
surface diffusion in random deposition model with varying 
extents of diffusion are performed and their results 
are interpreted in the light of the analytical 
calculations. Thus, a clearer understanding of the diffusion process
and the effect of diffusion length on surface roughness is obtained.
Systems with surface diffusion show nearly random 
deposition-like behaviour upto monolayer deposition. 
Their interface widths, in a logarithmic plot, are initially
linear, as in random deposition. With increase in the number of
layers, correlation effects between neighbouring
columns become dominant. The interface deviates from its initial
linear growth and eventually becomes saturated. An explanation for 
this behaviour is discussed and the point of departure 
from the linear form is estimated analytically.
\end{abstract}

\pacs{68.35.Fx, 68.35.Ct, 68.35.-p}

\keywords{Diffusion, interface roughness, correlated growth, combinatorial argument}
\maketitle

\section{Introduction}
Surface diffusion in random deposition model was introduced by
Family\cite{family86} to study the scaling properties of rough surfaces
under the effect of surface diffusion in $(1+1)$ dimensions.
In the model, particles are deposited from above onto a line of sites.
The particle arriving at any randomly chosen site, 
diffuses to the nearest neighbouring site that is 
lower than the selected site. The resulting surface is smooth 
compared to the surface in random deposition.
Diffusion to neighbouring sites introduces 
correlations among the neighbouring columns which eventually lead to
saturation of the surface roughness.

The roughness of the interface is quantitatively expressed
in terms of interface width $W$, which is a function of system size 
$L$ and the growth time $t$.
$W(L,t)$ is defined by the root-mean-square (rms) fluctuation
of the height of the interface. 
\beq\label{rms}
W(L,t) = \sqrt{ \frac{1}{L}\sum_{i=1}^L \left[h(i,t) -{\langle h(t) 
\rangle} \right]^2}
\eeq
Following Family-Vicsek\cite{famvic85} scaling relation,
$W(L,t)$ scales initially as $t^{\beta}$ for a fixed system size.
It saturates to a time-independent value $W_{sat}$,
beyond a model-dependent saturation time $t_{sat}$.
The saturated width scales with the system size $L$ as
$W_{sat} \sim L^{\alpha}$.

It was observed by Family, through simulations in one dimension, that 
the scaling exponents obtained are quite different from those 
obtained for random deposition (RD) \cite{barabasi}.
An interesting point, reported by Family, was that the surface properties
do not change with further increase in diffusion length.
However, no explanation was put forward for this independence.

Surface diffusion is found to be one of the most important 
factors affecting surface morphology in several practical 
applications.
For example, in molecular beam epitaxy (MBE), 
selective area metalorganic 
vapour phase epitaxy (SA-MOVPE) and chemical 
beam epitaxy of nanowires, the length of surface diffusion is significant in 
determining the growth rate and the mode of growth \cite{expt}. 
This diffusion length may be controlled by varying one or more 
experimental parameters such as the substrate temperature 
or introducing atomic or molecular hydrogen under pressure \cite{mori95}. 
An investigation is hence necessary to understand the role of 
surface diffusion in surface growth processes, and the nature 
of the dependence.

In Family's model, the extent of diffusion is confined
to nearest neighbours.
However, at any stage of deposition, there may be several sites with 
successively lower neighbours beyond the nearest.
At such sites, a priori it seems, 
roughness may be further reduced if diffusion 
is not limited to nearest neighbour alone. 

In this present work, we study in an analytical manner 
and also through numerical simulation, the significance of 
diffusion length in determining interface roughness.
Using simple combinatorial argument, the probability
with which a particle at any chosen site will diffuse
over a certain distance is derived, whence, for a given system
size and number of deposited particles, 
the probability of diffusion over different distances 
(nearest neighbour, next nearest neighbour etc.) can be 
estimated. 
The diffusion length with the maximum likelihood will be
the most dominant in determining the smoothness of
the surface.

To complement the analytical study,
numerical simulations of random deposition 
with surface diffusion are performed.
A newly arriving particle is allowed to diffuse over a
prescribed neighbourhood defined by a diffusion length $N_d$.
Starting with $N_d=1$, corresponding to Family's nearest 
neighbour diffusion, this diffusion length is varied over
the values $2,3,4,6, \dots 16$ and its effect on 
the dynamics of growth is observed. 
We shall refer to this extended Family's model
as random deposition with surface diffusion (RDSD).
The system sizes chosen, range from small sizes $24, \, 48, \, 
\cdots 384$, studied by Family, to larger sizes like 
$1024, \, 2048$ and $3072$.

Our analytical study shows that 
the nearest neighbour diffusion probability is 
most dominant, thereby explaining Family's observation 
that diffusion to farther neighbours does not affect 
surface properties.
The numerical simulations show a small change
in surface properties as the diffusion length is increased from $1, \,
2$ to $3$. Beyond $N_d=3$, the surface properties show no
further change, in accordance with our analytical results.

We address another related feature of RD and RDSD processes 
in the sub-monolayer region.
For RD, the logarithmic plot of the interface width with time
increases linearly. With the introduction of surface diffusion, 
the plot is nearly collinear with that for RD in the sub-monolayer 
region. A deviation from this linear form is evident as a 
complete layer of deposition is approached, much before the 
saturation of the interface width. This feature is discussed and 
a quantitative estimate of the time of this
departure from the straight line is made.

\section{The Model }
In Family's model of surface diffusion in random 
deposition\cite{family86}, a particle dropped 
in column $i$ sticks to the top
of the column $i$, $i+1$ or $i-1$ depending on which of the 
three columns has the smallest height. If the columns $i$, $i+1$ 
and $i-1$ have the same height, then the particle simply stays 
on top of column $i$ and does not diffuse. If the neighbouring 
columns $i+1$ and $i-1$ are of equal heights, 
lower than that of $i$, the particle is allowed 
to move to the top of 
either neighbour with equal probability. 

\begin{figure}[ht]
  \begin{center}
  {\includegraphics[width=\linewidth]{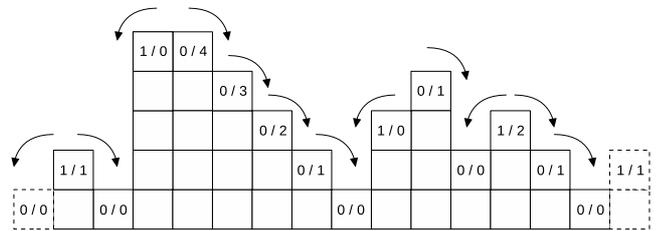}}
  \end{center}
  \caption{Schematic diagram for deposition rules in RDSD. 
The numbers (a/b) at the top of columns show maximum possible 
left diffusion (a) and right diffusion (b) at that site.}
  \label{rdsd}
\end{figure}

To extend the model to larger diffusion lengths, 
a particle arriving at a chosen site is 
allowed to diffuse to farther
neighbours on either side. As an illustration, for allowed diffusion 
length $N_d = 2$, a particle arriving at a chosen site $i$ is allowed 
to diffuse up to two neighbours, successively lower, on either side.
The particle first moves to the site lowest among $i$, $i+1$ and $i-1$.
If it has already moved to a neighbouring site, i.e., $i-1$ or
$i+1$, it then moves to the next neighbour, i.e., $i-2$ or
$i+2$ respectively, if that site is further lower. 
If in the first move $i-1$ and $i+1$ sites are equal
to each other and lower than site $i$, the movement to the left
or right is initiated by random choice. If however both 
sites $i-1$ and $i+1$ are equal to or higher than site $i$, no
movement takes place. 
Figure \ref{rdsd} illustrates the deposition rules for the model.

\section{Analytical Estimate of Diffusibility}
Let us consider a system of length $L$. The probability of selecting 
a random site among the $L$ available sites is $p = \frac{1}{L}$.
\begin{figure}[ht]
  \centering
  {\includegraphics[width=0.9\linewidth]{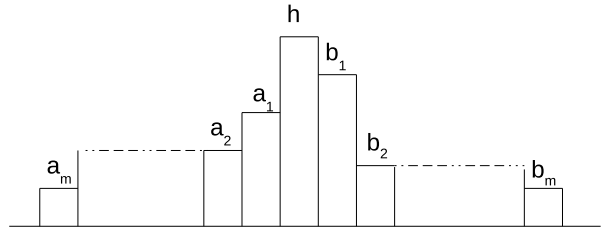}}
  \caption{A configuration where at least $m$ step diffusion
  on both sides is possible.}
  \label{mstepdif}
\end{figure}             
Let us first focus on the morphology of a small subsystem consisting
of the chosen site $i$ of height $h$ (where a new particle arrives) 
and its two nearest neighbours $(i-1)$ and $(i+1)$ with heights $a_1$
and $b_1$ respectively.
The configurations which allow at least one step
diffusion at the site $i$, say, to the left, must have the height 
$a_1$ of the ($i-1$)th column less than the height $h$ of the 
$i$th column, i.e., $a_1<h$. If $n$ layers, or a total of 
$nL$ particles are deposited, the probability that the 
chosen site has a lower left neighbour is given by  
\beq \label{oneleft}
\Pi_1=\sum_{h=1}^{nL} \; \sum_{a_1=0}^{nL-h} 
{^{nL}}C_{h} \;{^{nL-h}}C_{a_1}\; p^{h+a_1} \;(1-2p)^{nL-(h+a_1)} 
\eeq
The summations are subject to the restrictions 
$a_1 < h$ and $h+a_1 \le nL$.
Likewise, for diffusion to both left and right, at
the site $i$, by at least 
one step, the height of the left nearest neighbour
$a_1$ and that of the right nearest neighbour
$b_1$ should both be less than $h$. Then the probability
for at least one step diffusion to both sides 
is (See Figure \ref{mstepdif}),
\beqar \label{onelr}
\Pi_1^{L,R}&=&\sum_{h=1}^{nL} \; \sum_{a_1=0}^{nL-h} 
\;
\sum_{b_1=0}^{nL-h \substack{-a_1}} 
{^{nL}}C_{h} \;{^{nL-h}}C_{a_1} {^{nL-h-a_1}}C_{b_1} \nonumber \\
&& p^{h+a_1+b_1} \;(1-3p)^{nL-(h+a_1+b_1)} .
\eeqar
Here again, the summations are subject to the restrictions
$a_1 < h$, $b_1 < h$ and $h+a_1+b_1 \le nL$.
From the set of all configurations that allow at least one step
diffusion to the left at the chosen site, if one removes all 
those that allow diffusion to both sides by 
at least one step, one is simply
left with the configurations that allow diffusion only to
the left and not to the right. Hence, the probability of diffusing at 
least one step to the left with no diffusion to the right is given by
\bd
\Gamma_1^{(L)} = \Pi_1-\Pi_1^{(L,R)} \, .
\ed
From symmetry, we may write $\Gamma_1^{(R)}$, the probability 
of diffusion to the right, with no diffusion to the left as,
\bd
\Gamma_1^{(R)} = \Pi_1-\Pi_1^{(L,R)} \, .
\ed

The probability that a particle arriving at the chosen
site may diffuse to at least one neighbouring site on
either side is given by,
\beqar
\Lambda_1 &=& \Gamma_1^{(L)} + \Gamma_1^{(R)} + \Pi_1^{(L,R)} 
  \nonumber \\
& = & 2\Pi_1 - \Pi_1^{(L,R)}.
\eeqar
Extending the above arguments to $m$ diffusion steps,
the probability of diffusion by at least $m$ steps to
either side is given by,
\beq \label{lam}
\Lambda_m =  2\Pi_m - \Pi_m^{(L,R)} \;\;\;\;\;\;\;\;\; m = 1,2,\dots
\eeq
where, following Eq. (\ref{oneleft}) we may write
\beqa
\Pi_m &&= \sum_{h,\{a_k\}}
{ }^{nL}C_{h} \; {}^{nL-h}C_{a_1} \; \hfill \\
&& {}^{nL-(h+a_1+\dots+a_{k-1})}C_{a_k} \dots 
{}^{nL-(h+a_1+\dots+a_{m-1})} C_{a_m} \hfill \\ 
&&p^{h+a_1+\dots+a_m} 
[1-(m+1)p]^{nL-(h+a_1+\dots+a_m)}
\eeqa
where, the summations are over $h$
and every $a_k$, for $k=1,2,\dots,m$.
The restrictions on the summations are
$ 0\le a_m<a_{m-1}<\dots<a_1<h$
and
$h+a_1+\dots+a_m \le nL$.
Similarly, we may write for $\Pi_m^{(L,R)}$,
\beqa
\Pi_m^{(L,R)} &&= \sum_{h,\{a_k, b_k\}}
{}^{nL}C_{h} \; {}^{nL-h}C_{a_1} \; \hfill \\
&&{}^{nL-(h+a_1+\dots+a_{k-1})}C_{a_k} 
\dots {}^{nL-(h+a_1+\dots+a_{m-1})} C_{a_m} \\
&& {}^{nL-h}C_{b_1} \; {}^{nL-(h+b_1)}C_{b_2} 
\dots {}^{nL-(h+b_1+\dots+b_{m-1})} C_{b_m} \\
&& p^{h+a_1+\dots+a_m+b_1+\dots+b_m} \hfill\\
&& [1-(2m+1)p]^{nL-(h+a_1+\dots+a_m+b_1+\dots+b_m)}
\eeqa
where, the restrictions on the summations are
$ 0\le a_m<a_{m-1}<\dots<a_1<h$, 
$ 0\le b_m<b_{m-1}<\dots<b_1<h$ 
and $h+a_1+\dots+a_m + b_1 +\dots+b_m\le nL$.

The probability that a particle arriving at a selected site
will diffuse {\it{exactly}} $m$ steps is given by
\beq \label{finalpm}
\mathcal{P}_m = \Lambda_m - \Lambda_{m+1} \;\;\;\;\;\;\;\; m = 1, 2,\dots
\eeq
and the probability of zero or no diffusion is
\bd
\mathcal{P}_0 = 1 - \Lambda_1 .
\ed

\begin{table*}[!ht]
\centering
\begin{tabular}{c|c|c|c|c|c|c}
\hline
No. of Layers & 1 & 2 & 5 & 10 & 20 & 100 \\
\hline
$\mathcal{P}_0$& 0.5253 & 0.4533 & 0.4035&0.3810 & 0.3672&0.3478 \\
$\mathcal{P}_1$& 0.3882 & 0.4048  & 0.4050&0.4016 & 0.3970&0.3902 \\
$\mathcal{P}_2$& 0.08207 & 0.1252 & 0.1579&0.1728 & 0.1824&0.1918 \\
$\mathcal{P}_3$& 0.004444 &0.01578 & 0.02998&0.03855& 0.04499&0.05771  \\
$\mathcal{P}_4$& 0.00004930& 0.0009140& 0.003366&0.005471& 0.007388 &- \\
$\mathcal{P}_5$& 7.166 x 10$^{-8}$& 0.00002252 & 0.0002457&0.0005842& 0.0008818 &- \\
\hline
\end{tabular}
\caption{Probability $\mathcal{P}_m$ of a site 
with maximum diffusibility $m$ on either side after $n=1,2,5,10,20,100$
layers of particle deposited for system size $L=48$.}
\label{tab48}
\end{table*}
\begin{table*}[!ht]
\centering
\begin{tabular}{c|c|c|c|c|c|c}
\hline
No. of Layers & 1 & 2 & 5 &10 & 20 &100 \\
\hline
$\mathcal{P}_0$& 0.5279 & 0.4542 & 0.4172&0.3816 & 0.3673&0.3488  \\
$\mathcal{P}_1$& 0.3864 & 0.4046 & 0.4005&0.4011 & 0.3970&0.3900 \\
$\mathcal{P}_2$& 0.08105 & 0.1247 & 0.1514&0.18664 & 0.1824&0.1914\\
$\mathcal{P}_3$& 0.004516 & 0.01566 & 0.02762&0.02452 & 0.04497&0.05747 \\
$\mathcal{P}_4$& 0.00006098 & 0.0009081 & 0.002969&0.005475 & 0.007382&0.01064 \\
$\mathcal{P}_5$& 1.623 x 10$^{-7}$ & 0.00002336 & 0.0002037&0.0005439 &0.0008808 & -\\
\hline
\end{tabular}
\caption{Probability $\mathcal{P}_m$ of a site
with maximum diffusibility $m$ on either side after $n=1,2,5,10,20,100$
layers of particles deposited for system size $L=512$.}
\label{tab512}
\end{table*}


We have computed 
these probabilities for several system sizes.
In Table \ref{tab48} and Table \ref{tab512} 
the values are given for system sizes $L=48$ and $L=512$
respectively. We observe that the system sizes do not have any
marked influence on the probabilities.

\begin{figure}[ht]
  \begin{center}
  {\includegraphics[width=\linewidth]{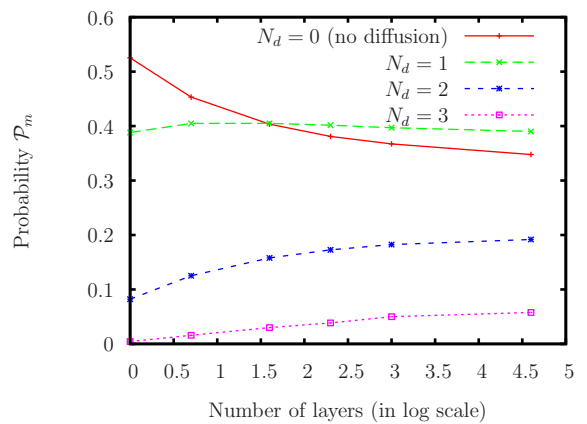}}
  \end{center}
  \caption{(Color Online) Probabilities of diffusion to zero, 
one and two steps plotted versus number of layers in log scale.}
  \label{diflen}
\end{figure}

\begin{figure}[ht]
  \begin{center}
  {\includegraphics[width=\linewidth]{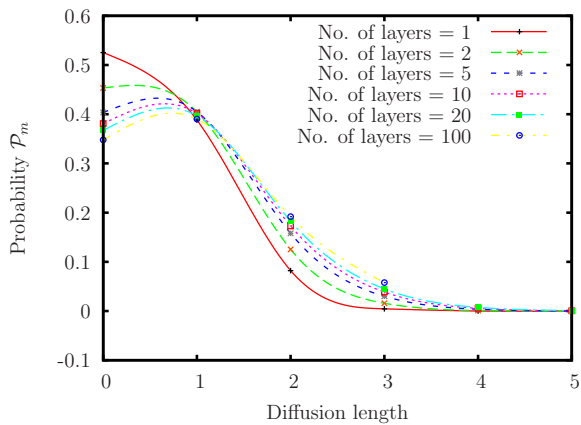}}
  \end{center}
  \caption{(Color Online) Diffusion probabilities versus diffusion length for
various layers. For more than two layers, the probabilities maximize
for nearest neighbour diffusion.} 
  \label{fermifit}
\end{figure}

A plot of $\mathcal{P}_m$ versus the number of layers in logarithmic
scale is shown in Fig \ref{diflen}. 
The probability that there is no diffusion
at all, steadily decreases as the number of deposited layers
increases. For finite diffusion lengths, all other probabilities
$\mathcal{P}_m$, $m=1,2,...$ 
increase as the number of layers increases.
After $10$ layers of deposition, the zero and one step diffusion
probabilities are the highest, followed by two and three step
diffusion probabilities. After $100$ layers, when the RDSD model
has attained saturation, the nearest neighbour 
diffusion probability dominates, followed by zero diffusion
probability as can be seen in 
Fig \ref{fermifit}. The probabilities for diffusion lengths $2$ and $3$
are significantly smaller in comparison, together accounting for less
than $1/3$ of the total possibility of diffusion.
The probabilities for higher diffusion lengths are
negligible.

In any given layer, beyond 
the first, the diffusion probability maximizes at diffusion length one.
Thus, one step diffusion has the most dominant effect in determining 
surface smoothness. 

\section{Numerical simulation study}
\begin{figure}[ht]
  \centering
  {\includegraphics[width=\linewidth]{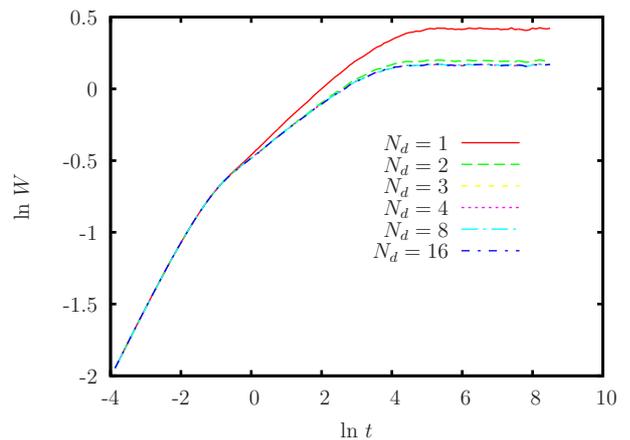}}
  \caption{(Color Online) $\ln W(t)$ vs $\ln t$ with 
      $L=48$ for different diffusion lengths.}
  \label{48all}
\end{figure}

Our simulations for nearest neighbour or one step diffusion include all system
sizes studied by Family, $L = 24,\dots, 384$ and beyond, up to $L=3072$.
Limiting our study to the system sizes studied by Family in $(1+1)$
dimension reproduces his results with great accuracy.
The study is then extended to larger diffusion lengths. 
For diffusion lengths $N_d=2,3,4,6,8\cdots$ 
we observe that a small change in the surface properties
is discernible up to $N_d = 3$. The surface roughness decreases
slightly, system saturates
at an earlier time and has a lower saturation width.
Beyond $N_d = 3$, the interface width versus time plots merge
(See Figure \ref{48all}).

It is important to note that in 
our simulations of RDSD,
two independent random number generators were used, 
one for selecting the site on the
growing surface and, another for depositing the particle with equal
probability when left and right diffusions are equally likely.
These two random number generators are completely independent and 
uncorrelated to each other and are generated by linear congruential 
method \cite{knuth,sedge}. 

\begin{figure}[ht]
  \centering
  {\includegraphics[width=\linewidth]{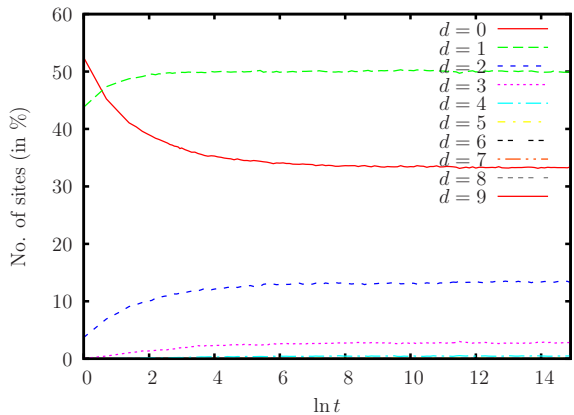}}
  \caption{(Color Online) Number of sites (in $\%$) with different
diffusibility versus time for Random Deposition Model.}
  \label{RDSat}
\end{figure}

To understand the results of simulation better,
we study the maximum possible diffusibility of sites 
at various stages of growth.
The maximum possible diffusibility for a given site
is the number of adjacent sites which are successively lower
in height (See Figure \ref{rdsd}).
For RD, we find that beyond sub monolayer region,
about $70\%$ of the sites allow diffusion to at least one step or
more, implying that an arriving particle
if allowed, will diffuse to one or more, successively 
lower nearest neighbours (See Figure \ref{RDSat}).
The plots are qualitatively similar to the plots
in Figure \ref{diflen}. The percentage of sites with a
certain diffusibility, say $m$, is a measure of the probability 
of diffusion by $m$ steps, and the number of layers deposited is
a measure of time. 
The diffusibility distribution is next
studied for Family's nearest neighbour model where
a newly arriving particle diffuses to a lowest nearest 
neighbour, even when farther lower sites are available.
The height profile is studied after saturation,
to ascertain the possibility of further diffusion at each site.
We observe that in the saturation region, for
about $9\%$ of the sites, further diffusion,
(beyond nearest neighbour) is possible.   
With the diffusion length extended to
$2$, it is observed that, for only $0.8\%$ of the sites, 
diffusion beyond two neighbours is possible.
With further increase in $N_d$,
the percentage of sites favouring diffusion greater than the maximum 
allowed limit set by
$N_d$, diminishes further. 
These observations agree with the diminishing 
probabilities $\mathcal{P}_m$ appearing in Eq. (\ref{finalpm}),
for increasing $m$. The analytically calculated probabilities
and the diffusibility distributions in the simulations explain the small 
change observed in the growth exponent and saturation width when 
the diffusion length is increased from $1, \, 2$ to $3$ steps. 
Further increase in diffusion length beyond $3$ has no discernible
effect on the surface properties obtained from simulations.

\begin{figure}[ht]
  \centering
  {\includegraphics[width=0.9\linewidth]{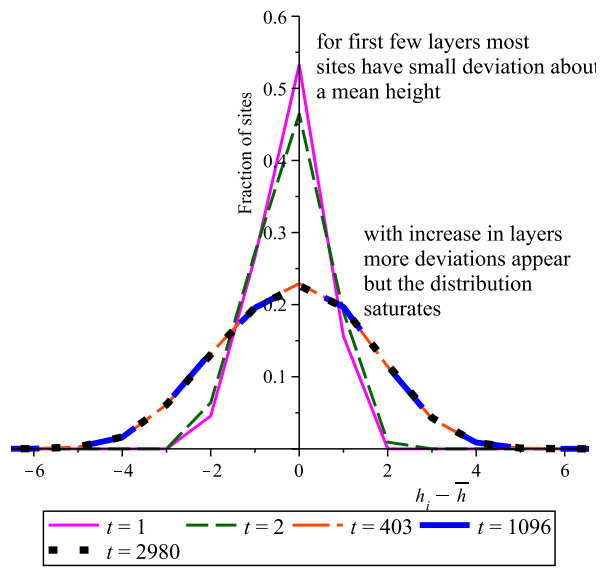}}
  \caption{(Color Online) Height distribution in simulation
of nearest neighbour diffusion model for system size 
$L=48$. The distributions are sharp and narrow 
for $ t < 403$ (growth region). The flatter, wider
curves for $t \ge 403$, overlap, showing saturation of the distribution.}
  \label{htgt}
\end{figure}

\begin{figure}[ht]
  \centering
  {\includegraphics[width=0.9\linewidth]{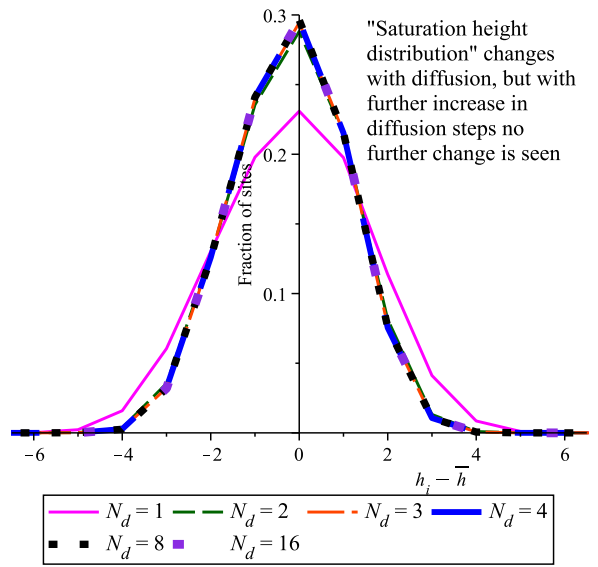}}
  \caption{(Color Online) Height distribution after saturation
in simulations of diffusion lengths $=2,3,4,8$ and $16$ 
for system size $L=48$.}
  \label{htsat}
\end{figure}

A study of the height profile, shows that in RD, the distribution 
of the heights of the sites about a mean height, becomes flatter 
and wider as more layers are deposited,
signifying an increase in interface roughness as time progresses.
When diffusion is allowed, this flattening
and widening process stops beyond a certain time, implying 
saturation of interface width. 
In Figure \ref{htgt}, this saturation 
of the height distribution is shown for diffusion length
$N_d = 1$ and system size $L=48$. 
The height distributions after saturation in RDSD models
depend on the diffusion length and are plotted
for $N_d = 1, 2, 3, 4, 8$ and $16$ in Figure \ref{htsat}.
It is noted that the saturated height distributions
do not change for diffusion length $N_d$ beyond $3$. 


In RDSD, the logarithmic graph of interface width versus
time, is initially, nearly collinear with the corresponding 
graph in RD; later it deviates from the linear form and
eventually saturates.
We estimate the quantitative nature of this
deviation. 
Let us consider a random deposition process with
system size $L$, number of particles deposited $N$ and
$N < L $ defines sub-monolayer deposition. 
The probability of a site being occupied is $\frac{N}{L}$
and that being empty is $1-\frac{N}{L}$.
In the sub-monolayer random deposition, a chosen site may 
allow diffusion provided either one or both its nearest 
neighbours are empty.
Thus, the probability of finding a site
where diffusion may be possible is given by
$\big(\frac{N}{L}\big) \big(\frac{N}{L}\big) \big(1-\frac{N}{L}\big) 
+ \big(\frac{N}{L}\big)\big(1-\frac{N}{L}\big) \big(\frac{N}{L}\big) 
+\big(\frac{N}{L}\big) \big(1-\frac{N}{L}\big) \big(1-\frac{N}{L}\big)$ 
which is equal to $\frac{N}{L}\big(1-\frac{N}{L}\big)
\big(1+\frac{N}{L}\big)$. 
\begin{figure}[ht]
  \centering
  {\includegraphics[width=\linewidth]{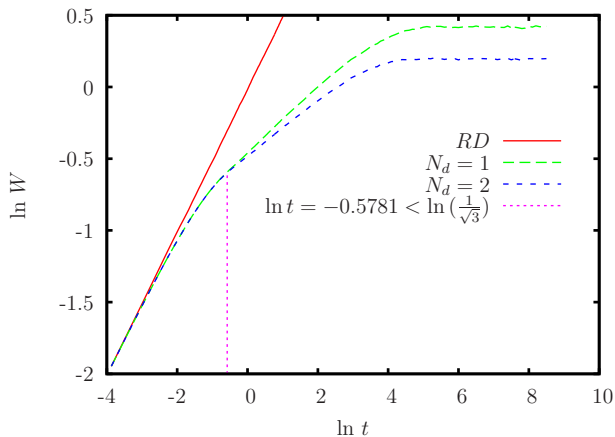}}
  \caption{(Color Online) $\ln W(t)$ vs $\ln t$ showing deviation from
    the straight line nature.}
  \label{value}
\end{figure}

The effects of diffusion are most noticeable when this probability
function is maximum. Denoting $\frac{N}{L}=t$, it is easily
verified that the maximum for the above
probability function $f(t)= t \,(1-t^2)$ occurs when 
$t = \frac{1}{\sqrt{3}}$. 
Thus for $t \simeq \frac{1}{\sqrt{3}} $,
$f_{max} \simeq 0.4$, which implies no more than $40\%$ of the sites
in the sub-monolayer in RD, allow diffusion. 
In an actual process, which involves diffusion, we conjecture that
the above probability function will reach the maximum value
at an even lower value of $t$.
Deviation from random behaviour, in an actual process,
is visible for $t < \frac{1}{\sqrt{3}}$ and 
the deviation will become significantly large
for $t \geq \frac{1}{\sqrt{3}}$.

In Figure \ref{value}, the results of 
an actual simulation for $L=48 , N_d=1,2$ and RD are shown where
departure from random behaviour is evident for $\ln t \ge \ln \, 
\frac{1}{\sqrt{3}}$.
\section{Discussion and Conclusion}
We have shown by means of analytical calculation as well
as numerical simulation that in RD, 
appearance of long staircase-like formations 
with successive sites having ascending or 
descending heights, even after several layers of deposition,
is improbable. After a few successive 
descending sites ($1$ or $2$), 
one reaches the local minimum of a staircase.
Thus, even in the case of RD, maximum diffusibility of sites is limited.
However, the differences in the heights of unequal nearest 
neighbours increase without limit. This is in agreement with the
fact that roughness of the interface increases as $t^{1/2}$,
even though the maximum diffusibility at any site is limited.
It is interesting to note that the 
number of sites with different diffusibility
approaches stable values as time progresses (See Figures \ref{diflen}
and \ref{RDSat}).

The study of the effect of diffusion can be extended to systems
that allow diffusion through flat regions or even through
barriers. They may represent continuation of rolling of classical
particles through flat regions due to inertia or quantum barrier
penetration, e.g., chemical deposition process even for 
reactants separated by potential barrier.
Further investigation pertaining to the above processes 
are in progress.

The authors wish to acknowledge the computational facility extended by 
Center for Mobile Computing and Communication
of Jadavpur University.
One of the authors (BM)
also wishes to acknowledge the financial assistance provided by
the West Bengal State Departmental Fellowship (India).

\end{document}